# Assessing the optimal contributions of renewables and carbon capture and storage toward carbon neutrality by 2050

Dinh Hoa Nguyen [a,b,*], Andrew Chapman [a], Takeshi Tsuji [a]

[a] International Institute for Carbon-Neutral Energy Research (WPI-I2CNER),

[b] Institute of Mathematics for Industry (IMI),

Kyushu University, Fukuoka 819-0395, Japan

Emails: hoa.nd@i2cner.kyushu-u.ac.jp, chapman@i2cner.kyushu-u.ac.jp, tsuji@mine.kyushu-u.ac.jp

**Abstract:** Building on the carbon reduction targets agreed in the Paris Agreements, many nations have renewed their efforts toward achieving carbon neutrality by the year 2050. In line with this ambitious goal, nations are seeking to understand the appropriate combination of technologies which will enable the required reductions in such a way that they are appealing to investors. Around the globe, solar and wind power lead in terms of renewable energy deployment, while carbon capture and storage (CCS) is scaling up toward making a significant contribution to deep carbon cuts.

Using Japan as a case study nation, this research proposes a linear optimization modeling approach to identify the potential contributions of renewables and CCS toward maximizing carbon reduction and identifying their economic merits over time. Results identify that the combination of these three technologies could enable a carbon dioxide emission reduction of between 55 and 67 percent in the energy sector by 2050 depending on resilience levels and CCS deployment regimes. Further reductions are likely to emerge with increased carbon pricing over time.

[*] Corresponding author

The findings provide insights for energy system design, energy policy making and investment in carbon reducing technologies which underpin significant carbon reductions, while identifying potential regional social co-benefits.

## 1. Introduction

As nations around the world contend with ambitious carbon reduction goals, predominantly derived from the Paris agreements [1], Japan, under former Prime Minister Suga, has declared that it will become carbon neutral by 2050 [2]. As part of this declaration, alongside innovations such as renewable energy (RE), the management of carbon is also expected to play a role. Carbon capture and storage (CCS) represents a technology which could assist in rapidly reducing the carbon dioxide emissions, particularly those from electricity and heat, i.e., from fossil fuel power stations, responsible for some 610 million tons (Mt) or ~52% of $CO_2$ emissions each year in Japan [3]. The role and timeline for commercialization of CCS is discussed in the Basic Energy Plan of Japan [4], and the processes of capture, transportation, injection and storage in Japan are under investigation along with studies on suitable sites for storage. At the global scale, CCS has also been identified as one of the key pillars to achieve net-zero $CO_2$ emissions by 2050 [5].

CCS, however, is one approach among many for carbon reduction. These include RE, nuclear power, energy efficiency and forestry, among others. For example, carbon capture and utilization (CCU) has the potential to take $CO_2$ from the atmosphere and convert it into useful products through energy provided by renewables, potentially engendering a carbon negative outcome [6]. There is also the potential to use renewable energy as a heat source to convert biomass to create low-carbon hydrogen to offset the use of fossil fuels [7]. Recently, the cost of RE has decreased significantly [8], identifying the potential for replacement of some fossil fuels currently used for power generation which is responsible for a significant share of carbon emissions. Therefore, research is required to determine the appropriate contribution from each carbon reducing technology or approach, according to their scale of potential contribution and economic merits. Taking into account the unique Japanese situation, where all fossil fuels are imported, the role, scale and cost of alternative approaches to energy generation and carbon management are required.

Toward achieving carbon neutrality, along with a suite of technologies, a number of policy approaches exist, including feed in tariffs (FITs), carbon taxes, renewable portfolio standards and carbon trading, to name a few. Each has its own benefits and drawbacks, and can tend to favor certain technologies, as has been the case for solar power under the FIT in Japan [9].

The aim of this research is to uncover the optimal combination of technologies to achieve carbon neutrality at the best cost, i.e., empowering the market to choose the best technologies based on their environmental and economic merits conscious of varying energy policy approaches. Further, we seek to explore the policy settings, including the carbon price required to stimulate different carbon reducing technology deployments in Japan, to the target year of 2050, cognizant of recent Japanese carbon reduction ambitions.

This paper is structured as follows. Section 2 investigates previous research contributions in this area, identifying the gaps filled by this research. Section 3 details the methodology used to investigate potential future combinations of renewable energy and CCS to best contribute to Japanese energy goals. Section 4 describes the results of our linear optimization model across multiple future scenarios. Section 5 discusses the findings, including technological, environmental, and economic merits along with policy implications. Finally, Section 6 describes the conclusions, limitations and future directions of this research.

## 2. Background and Literature Review

This study builds on a body of work which has investigated the potential for national and regional emission trading and carbon reducing technology combinations. Previous modeling efforts have considered the suite of existing and emerging technologies required to meet carbon reduction goals in Japan [10]. Some of these models consider hydrogen as a key technology, while still recognizing the strong role required of CCS in achieving decarbonization [11]. In regional modeling efforts, the role of electrification, the need for energy carriers such as hydrogen and the role of CCS in decarbonizing fossil fuels and some industrial processes is also recognized [12]. Most recently, Nguyen et al., detailed an emission trading system (ETS) model which incorporated the technologies of wind and solar power in Japan, to maximize carbon reductions, cognizant of energy system resilience and best cost [13]. This work identified that an ETS can increase the amount of renewable energy deployed overall, however, requiring a resilient energy system reduces overall deployment while increasing energy system cost. With regard to return on investment, it was also clarified that a carbon price approaching $100 USD is required to keep payback periods under 20 years [13] for investors in renewable energy deployment.

As many nations move toward carbon neutrality, researchers are investigating the potential of ETS to increase renewable deployment, along with complementary carbon reducing technologies (including CCS) to meet national carbon reduction goals. For example, it was identified that for China to achieve its Paris Agreement targets, that an ETS creates a potential least cost system. A carbon price above $40 per ton of $CO_2$ ($tCO_2$) was found to be conducive to wind power and coal fired power with CCS, however, for a wide-scale deployment of CCS, prices above $100/$tCO_2$ are required to achieve national carbon goals [14]. These findings are complemented by Zhou et al., who found that in order to stimulate CCS deployment for combined heat and power plants, different carbon prices engender differing CCS retrofitting timelines, i.e., the year 2033 for a carbon price of $14.5 USD/ $tCO_2$, 2030 at $20.7, and as early as 2025 at prices above $23.4/$tCO_2$ [15]. These prices are much lower than European Union (EU) ETS prices. A study on the policies which are conducive to solar photovoltaic (PV) deployment and CCS deployment in China, namely power tariffs and an ETS found that low carbon prices disadvantaged CCS compared to PV, and in order to engender further CCS

deployment, power tariffs would need to be rebalanced, or carbon prices significantly increased [16].

For the EU, in order to achieve carbon neutrality by 2050, a significant tightening of the ETS regulations was identified by Pietzcker et al., who showed that with more ambitious targets (i.e., -63% $CO_2$ by 2030) and a carbon price of about 129 Euros/t$CO_2$, RE could make up to 74% of electricity by 2030, and fossil fuel-based generation could be phased out approximately 15 years earlier than under previous conditions. Under ambitious energy system transition pathways, CCS only plays a small role, and overall, energy system costs increase by a moderate 5% compared to 2020 levels, despite a tripling of carbon prices compared to the reference targets [17]. When investigating multiple modeling approaches to the European Union ETS, Ruhnau et al., found that different models can give different results, with a carbon price of 27 Euros/tCO2 in 2030 effects a decrease in emissions in the range of 36-57%, while higher prices of 57 and 87 Euros/t$CO_2$ yield reductions in the range of 45-75% and 52-80% respectively. They described the variance in emission reductions due to a number of factors including market driven decommissioning of fossil fueled power generation, fuel switch impacts captured by dispatch type models, and the consideration of market based investments in renewables [18].

For Japan, the case study nation of this research, a nationwide ETS is yet to be successfully conducted, with test cases only occurring in Tokyo and Saitama, engendering relatively low carbon prices of between $2 and $12/t$CO_2$ and limited trading between entities [13]. Recent evaluation of the Tokyo ETS identified that participant's carbon reductions were due to electricity price increases as a result of the Great East Japan Earthquake in 2011, and the ETS, in approximately equal parts [19].

Another issue, which is central to the deployment of an ETS, is how it affects the sharing of costs and benefits, or if certain regions benefit, at the expense of others. In China for example, the pilot ETS system was found to reduce urban-rural income inequality, however the effect was most pronounced in regions with relatively high CO2 emissions and per capita GDP [20], meaning that lower emitting and per capita GDP regions did not realize the same benefits. For the EU on the other hand, carbon taxes and ETS as strategies toward carbon neutrality were found to comparatively disadvantage lower income households, increasing the risk of energy poverty [21]. This risk may be alleviated through a redistribution of ETS revenues toward these at-risk households. Under a global evaluation of carbon pricing regimes, Chepeliev et al., found that inter-regional inequality moderately increases, while intra-region inequality is decreased [22]. Overall, lower economic growth leads to a slightly higher incidence of global poverty levels, however, carbon pricing regimes also allow for burden shifting to higher income households and a change in pricing outcomes for other necessities causing positive effects in some nations. Overall, there is agreement among researchers that redistribution of carbon pricing regime profits in a progressive manner may alleviate some of the negative outcomes [22,23]. In terms of redistribution, progressive distributional outcomes are suggested to be more likely for lower income nations, and also for regimes which consider indirect effects and consumer spending patterns [24].

In our previous investigation of Japan, prior to the incorporation of CCS as a complementary carbon reducing measure, we identified that different carbon target and energy system settings engendered different RE deployment outcomes for the 47 prefectures of Japan, with a resilience constraint increasing overall participation and benefit sharing [13].

This study seeks to evaluate an ETS in Japan which can incorporate CCS as a complementary carbon reducing approach, cognizant of geographic and cost limitations. The novelty of this research is the capability of our model to pinpoint where, when, and in what quantity renewables and CCS should be deployed to achieve both carbon reduction and energy system resilience targets. Further, we identify the optimal $CO_2$ transportation route, forming the basis of a peer-to-peer carbon emission trading system between prefectures. Based on the technological, economic and environmental findings, this study identifies the social outcomes of these deployment regimes over time, identifying the sharing of costs and benefits at the prefectural level. Our model is not limited to the case study nation of Japan and is adaptable to various nations and regions where appropriate data are available.

## 3. Methodology

This section describes our proposed modeling approach for assessing the maximum carbon emission reduction potential in a large geographical area (e.g., nation, continent, inter-continent, etc.) composed of smaller regions, each of which have potential RE sources and some of which have physical CCS deployment potential, while others do not. Hence, the proposed model is well fitted to reality, where physical CCS sites are available only at some specific regional locations. Moreover, the proposed model is helpful for analyzing the potential carbon emission trading between smaller regions, depending on the existence of physical CCS storage sites.

The proposed model is presented in the form of a linear programming problem over a defined time period, whose objective is to minimize investment costs while maximizing benefits obtained by regions as a result of carbon emission trading and RE-based electricity trading with the national grid. Model inputs consists of a yearly carbon emission cap for the whole large geographical area, RE and CCS potentials of each of the regions, unit costs for RE and CCS deployments, unit transportation cost for CCS, carbon emission trading prices, geographical distances between smaller regions, FIT electricity prices for RE-based generation, and RE conversion factors. On the other hand, model outputs consist of the amounts of CCS and each type of RE source deployed in each discrete region in each year over the analyzed time period. A schematic of the proposed model is depicted in Figure 1. Details of the proposed model are given below.

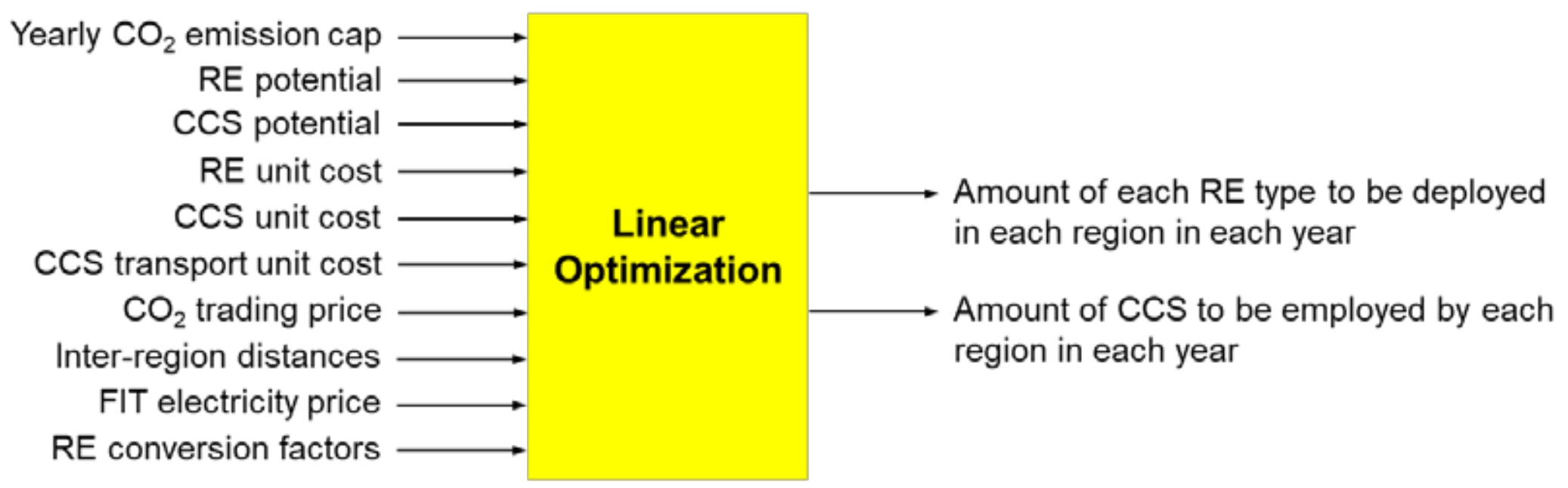

**Figure 1**. Illustration of the proposed linear optimization model.

### 3.1 The Proposed Optimization Model

The variables used in the proposed model and their definitions are detailed in Table 1.

**Table 1**. Variables and Functions used in the proposed model

| Variable | Definition | Meaning |
|---|---|---|
| $n$ | | Number of prefectures |
| $i$ | | Subscript for prefecture index |
| $t$ | | Time index [year] |
| $T$ | | Number of years |
| $k$ | | RE index |
| $K$ | | Number of RE types |
| $V_s$ | | Set of prefectures with physical $CO_2$ storage capability (CCS selling prefectures) |
| $V_b$ | | Set of prefectures having no physical $CO_2$ storage capability (CCS buying prefectures) |
| $RE_{i,k}(t)$ | Capacity of RE type $k$ in prefecture $i$ [GW] | The amount of RE type $k$ in prefecture $i$ to be deployed at year $t$ |
| $g_{i,k}\ (t)$ | Conversion ratio from RE type $k$ in prefecture $i$ to $CO_2$ emission [t/GW] | Showing how much $CO_2$ emission can be reduced by installing 1 GW of RE type $k$ in prefecture $i$ |
| $C_i(t)$ | $CO_2$ emission [thousand ton/year] | Showing the $CO_2$ emission in prefecture $i$ at year $t$ obtained by emission trading system |
| $rp_{i,k}(t)$ | Unit cost of RE type $k$ in prefecture $i$ [Y/GW] | The cost for deployment of 1 GW of RE type $k$ in prefecture $i$ |
| $cap(t)$ | $CO_2$ cap [ton] | The cap on $CO_2$ emission in Japan at year $t$, set by the emission reduction target |
| $RE_{i,k}^{\max}(t)$ | Maximum capacity of RE type $k$ in prefecture $i$ [GW] | Maximum remaining potential of RE type $k$ in prefecture $i$ at year $t$ |
| $cp(t)$ | $CO_2$ price [Y/t] | Price for a ton of $CO_2$ emission to be traded |
| $h_{i,k}$ | Conversion factor of RE type $k$ in prefecture $i$ from GW to GWh | Showing how many GWh is obtained by deploying 1 GW of RE type $k$ in prefecture $i$ |
| $sp_k(t)$ | Feed-in-tariff electricity price for generation from RE type $k$ [Y/GWh] | Price for 1 WGh electricity generated by RE type $k$ |
| $CCS_i^s(t)$ | Captured $CO_2$ emission [thousand ton/year] | Showing the amount of $CO_2$ emission captured by prefecture $i$ with physical $CO_2$ storage capability, at year $t$ |
| $CCS_{j,i}^b(t)$ | Traded $CO_2$ emission for CCS [thousand ton/year] | Showing the amount of $CO_2$ emission virtually captured by prefecture $j$ having no physical $CO_2$ storage capability, but bought |
| | | from prefecture $i$ with physical $CO_2$ storage capability, at year $t$ |
| $ccsp(t)$ | $CO_2$ capture price [Y/t] | Unit price for a ton of $CO_2$ emission to be captured and stored by CCS deployment |
| $gt(t)$ | Unit transportation cost [Y/(t x km)] | Unit price for ground transport of a ton of $CO_2$ emission to storage sites incurred for prefecture $i$ at year $t$ |
| $d_{ij}(t)$ | Distance [km] | Distance between prefecture $j$ having no storage site and prefecture $i$ with storage capability for storing $CO_2$ at year $t$ |
| $CCS_i^{max}(t)$ | Potential of captured and stored $CO_2$ emission [thousand ton] | Showing the maximum amount of potential $CO_2$ emission captured and stored by CCS deployment at prefecture $i$ at year $t$ |
| **Function** | **Definition** | **Meaning** |
| $IV_{i,k}(t)$ | Investment cost of RE type $k$ in prefecture $i$ [Y] | Cost for installing the traded amount of RE type $k$ in prefecture $i$ for emission trading with other prefectures |
| $EF_{i,k}(t)$ | Economic function of $CO_2$ emission in prefecture $i$ [Y] | How much prefecture $i$ gains from emission trading system for deploying the traded amount of RE type $k$ |
| $PF_{i,k}(t)$ | Profit function of selling electricity from RE type $k$ generation in prefecture $i$ [Y] | How much prefecture $i$ gains by selling electricity from the deployed amount of RE type $k$ |
| $IC_i(t)$ | Investment cost of CCS by prefecture $i$ [Y] | Cost incurred for prefecture $i$ for storing $CO_2$ at specific CCS sites |
| $TC_i(t)$ | CCS transportation cost for prefecture $i$ [Y] | Cost incurred for prefecture $i$ for transporting $CO_2$ to specific CCS sites |
| $SC_i(t)$ | CCS storage cost for prefecture $i$ [Y] | Cost incurred for prefecture $i$ for storing $CO_2$ at specific CCS sites |

The individual components included in the objective function of the proposed linear optimization model are described as follows:

First, the $CO_2$ emissions of each prefecture, which are used for trade with other prefectures via RE installation and CCS, is computed by:

$$C_i(t) = C_i(t-1) - \sum_{k=1}^{K} g_{i,k}(t) * RE_{i,k}(t) - CCS_i(t) \tag{1}$$

Second, the investment cost of RE technologies is:

$$IV_{i,k}(t) = rp_{i,k}(t) * RE_{i,k}(t) \tag{2}$$

Third, the economic function for $CO_2$ emissions obtained with RE deployment is:

$$EF_{i,k}(t) = cp(t) * g_{i,k}(t) * RE_{i,k}(t) \tag{3}$$

Fourth, the profit function of selling electricity from renewable generation is computed by:

$$PF_{i,k}(t) = sp_k(t) * h_{i,k} * \sum_{\tau=1}^{t} RE_{i,k}(\tau) \tag{4}$$

Fifth, the investment cost of CCS is:

$$IC_i(t) = ccsp(t) * CCS_i(t), i \in V_s \tag{5}$$

For the prefectures with no physical injection sites, i.e., $i \in V_b$, their CCS investment costs are zero.

Sixth, the cost for transporting $CO_2$ to the prefectures (regions) which have physical injection sites is computed by:

$$TC_j(t) = gt(t) * \sum_{i \in V_s} CCS_{j,i}^{b}(t) * d_{ji}(t), j \in V_b \tag{6}$$

This transportation cost will be paid to third parties who actually conduct such $CO_2$ transport. For the prefectures having physical injection sites, their $CO_2$ transportation costs $TC_i(t)$, $i \in V_s$, are obviously zero.

Seventh, the cost for $CO_2$ storage at physical CCS sites needs to be paid by CCS buying prefectures and transferred to CCS selling prefectures, computed thus:

$$SC_j(t) = cp(t) * CCS_j(t) = cp(t) * \sum_{i \in V_s} CCS_{j,i}^{b}(t)\,, j \in V_b \qquad (7)$$

Considering all of the components introduced above, the overall objective function in the proposed linear programming model is provided in (8). The constraints in this optimization model are as follows. The yearly cap for $CO_2$ emission in the whole considering geographical area is described by the inequality in (9). The CCS buying constraint for regions having no physical CCS storage sites is given in (10). To account for the limit on $CO_2$ storage in different regions, (11) is introduced. The yearly limited potential of each RE type in each region is shown in (12). Next, a resilience constraint placed upon the RE mix, i.e., the ratio of different types of RE to be deployed to contribute toward a stable power supply (i.e., the most available RE, engender through a ratio of 31% solar and 69% wind, detailed in [25]), is presented via the inequality in (13), where $\alpha_k > 0$ are given parameters to represent yearly limits on the total amounts of specific RE types to be installed based on a given energy policy.

Finally, the proposed linear optimization model is presented below:

minimize

$$\sum_{t=1}^{T}\sum_{i=1}^{n}\sum_{k=1}^{K}\left[IV_{i,k}(t) + IC_i(t) + TC_i(t) + SC_i(t) - PF_{i,k}(t) - EF_{i,k}(t)\right] \qquad (8)$$

subject to:

$$\sum_{i=1}^{n} C_i(t) \le cap(t) \qquad (9)$$

$$CCS_{j,i}^{b}(t) \geq 0, j \in V_b, i \in V_s \tag{10}$$

$$0 \leq \sum_{j \in V_b} CCS_{j,i}^{b}(t) + CCS_{i}^{s}(t) \leq CCS_{i}^{max}(t), i \in V_s \tag{11}$$

$$0 \leq RE_{i,k}(t) \leq RE_{i,k}^{\max}(t) \tag{12}$$

$$0 \leq \sum_{i=1}^{n} RE_{i,k}(t) \leq \alpha_k \sum_{k=1}^{K} \sum_{i=1}^{n} RE_{i,k}(t) \tag{13}$$

Solving the linear optimization problem (8) with constraints (9)–(13) will identify the following:

(i) the maximum amount of carbon emission reductions which can be achieved by 2050; and,

(ii) the answers to the key questions of where, when, and how much solar, wind, and CCS should be deployed during the period 2018–2050.

There are three sources of uncertainty in the proposed model. The first source of uncertainty comes from the $CO_2$ price for trading, i.e., $cp(t)$. The second source of uncertainty is due to the uncertainty on the unit cost of CCS deployment, i.e., $ccsp(t)$. And the last source of uncertainty is on the CCS unit transportation cost to the CCS storage sites, i.e., $gt(t)$. Therefore, model outcomes will be significantly influenced depending on how these costs vary. On the one hand this increases the difficulty of model operation, but also introduces the prospect of a range of resultant scenarios as settings are varied according to energy policy settings. For example, we can expect that for a fixed $CO_2$ price, any change in CCS unit price will influence the amount of $CO_2$ emissions captured by CCS each year. Further, any variation of $CO_2$ price will impart influence on the above amount of $CO_2$ emissions captured by CCS. It is the contrast of this range of potential scenarios and results which will bring insights for policy implications to the fore.

To facilitate the analysis of the proposed model and to compare outcomes with a previous study [13] in which CCS was not considered, Japan is employed as a case study nation hereafter. As such, each smaller region in the proposed model is associated with a prefecture in Japan (47 prefectures in total). The time period considered is from 2018 to 2050 to achieve greatest possible contribution toward the set carbon emission reduction goal of carbon neutrality. Data for model inputs are taken from the following sources, which are the same as that in [13] for the comparison purpose.

1. RE deployment potential (current and economically feasible future deployment; [21,22])
2. RE unit investment costs [8]
3. RE technology learning curves [27]
4. RE conversion factors (wind speeds [28] and solar insolation [29,30])
5. Current carbon emission profiles for each energy generation region (sourced from generator annual reports)
6. RE lifecycle GHG intensities [31]
7. CCS sites and potential for Japanese prefectures [32]

In addition to the calculation of RE deployment, investment cost and ETS revenues, the value of electricity generated by each RE source is also calculated using current and future projected FITs for each RE type to the year 2020 for solar PV (10 years for small scale and 20 years for large scale contracts greater than 50kW capacity), and the year 2020 for wind (20 year contracts), reverting to 8 yen per kWh post 2020 (in line with current projections and expected end of feed in tariff payment levels [33]). Solar panel and wind turbine replacement times are set conservatively at 20 years [34].

Furthermore, consideration of RE deployment is limited to solar and onshore wind, i.e., $k = 1$ corresponds to solar energy, while $k = 2$ represents the index for wind energy.

Figure 2 shows how the proposed optimization problem is solved to derive the optimal deployment quantities of solar, wind, and CCS over the considered time period.

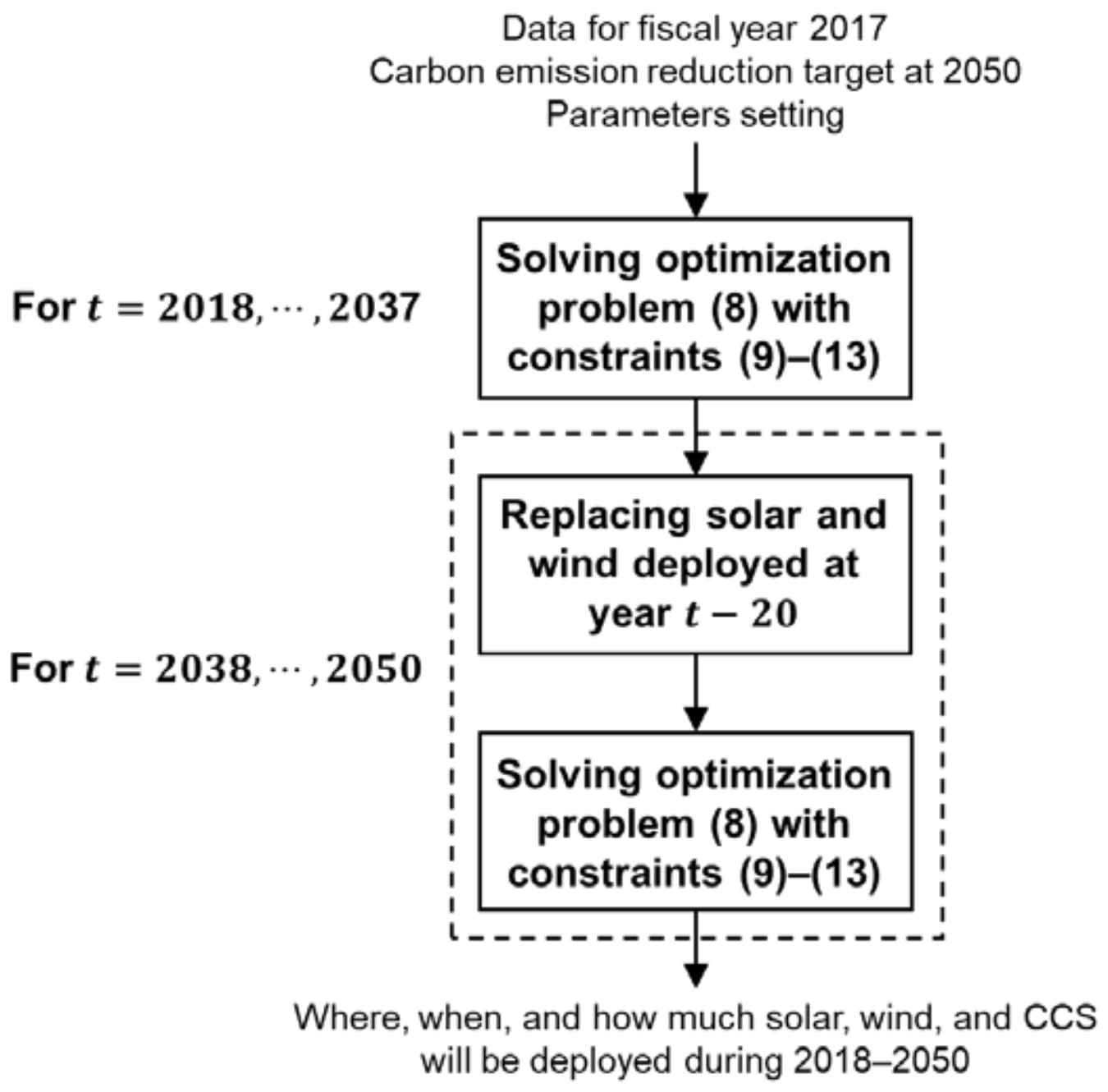

**Figure 2**. Solving steps in the proposed linear optimization model.

## 4. Results

In this section, results for four future energy system scenarios are simulated and detailed for the proposed linear optimization model as follows:

- Scenario 1: The physical limits of CCS storage capacity determined in Eq. (11) is equally divided for the years in the period 2018-2050. The power mix resilience constraint in Eq. (13) is not considered.
- Scenario 2: Similar to Scenario 1, but the power mix resilience constraint in Eq. (13) is taken into account.
- Scenario 3: No yearly constraint for the physical limits of CCS storage capabilities, only the total amount of CCS storage capacity as in Eq. (11) is considered for the period 2018-2050. The power mix resilience constraint in Eq. (13) is not considered.
- Scenario 4: Similar to Scenario 3, but the power mix resilience constraint in Eq. (13) is considered.

The first important outcome from simulation results is that consistently deploying CCS incrementally over the investigated time period is much better than rushing to deploy in the early years, due to:

- A much higher amount of overall $CO_2$ emission reductions achieved, and
- An earlier start for and larger distribution of renewable deployment over prefectures and time.

The second important outcome from simulation results is that CCS trading does not occur, even though the carbon and CCS prices are varied in the model. Instead, renewables deployment and local CCS injection at prefectures with physical storage sites is preferred throughout the investigated period. This can be explained via the economic viability of RE and CCS. More specifically, RE deployment in our model is more profitable than that of CCS, due to the accumulated profits obtained, due to RE-derived electricity sold back to the grid. Since our proposed model seeks to maximize profits while minimizing investment costs, local RE installation and CCS deployment are preferred to CCS trading between prefectures.

Simulation results for each scenario are detailed with the carbon price, the CCS unit cost, and the CCS transportation cost fixed at 10,000 Y/ton, 10,000 Y/ton, and 8.1739 Y/ton per km, respectively. The amount of $CO_2$ emissions for the whole of Japan in 2017 was approximately 1.25 billion tons [35].

***Scenario 1:*** The maximum amount of $CO_2$ which can be offset is 67%, which is much greater than the maximum of 42% $CO_2$ reduction achieved in a previously investigated scenario without the use of CCS [13]. This number is equivalent to approximately 834.24 million tons of $CO_2$ offset by 2050, which clearly demonstrates the role of CCS in massively reducing carbon emissions overall.

The results for deployment of solar, wind, and CCS are shown in Figures 3–4, respectively. For clarity of representation, only prefectures with high deployments of solar, wind, and CCS are shown in these figures and all subsequent figures.

In this scenario, wind deployment starts earlier than solar installation, whereas the deployment of CCS is undertaken in equal yearly increments throughout the considered period in each prefecture with physical storage capability.

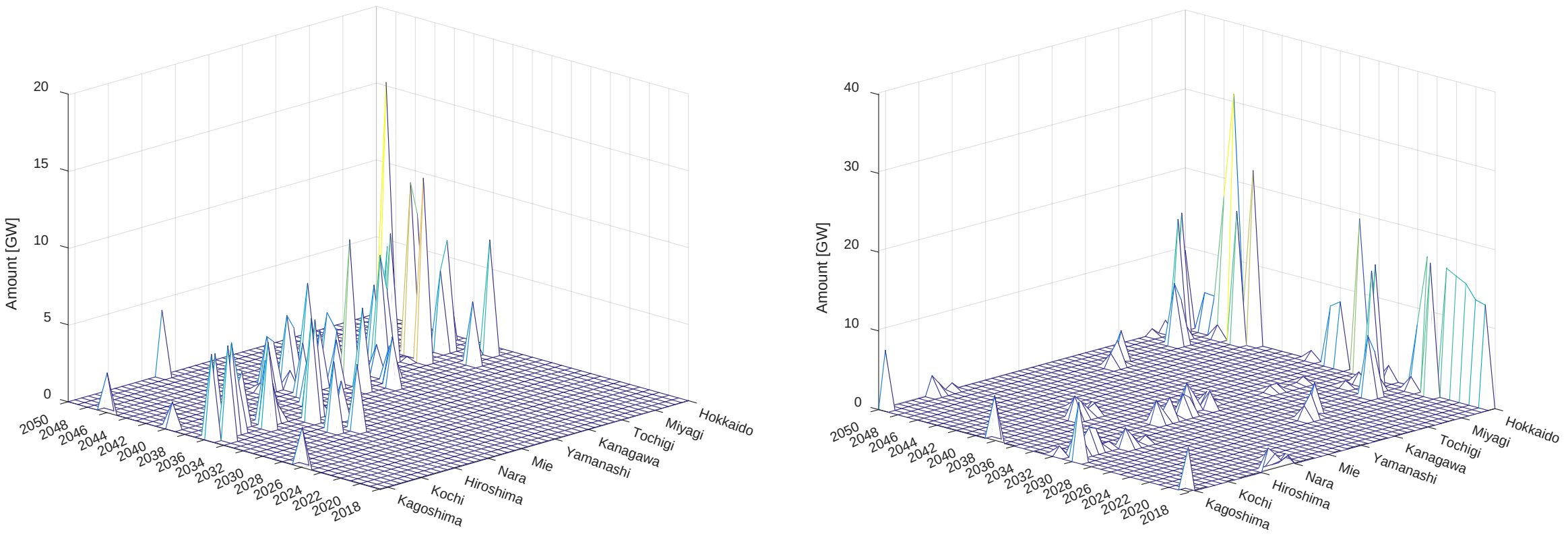

**Figure 3**. Yearly solar (left) and wind (right) deployment obtained from the proposed model between 2018–2050 for Scenario 1 enabling a 67% $CO_2$ reduction.

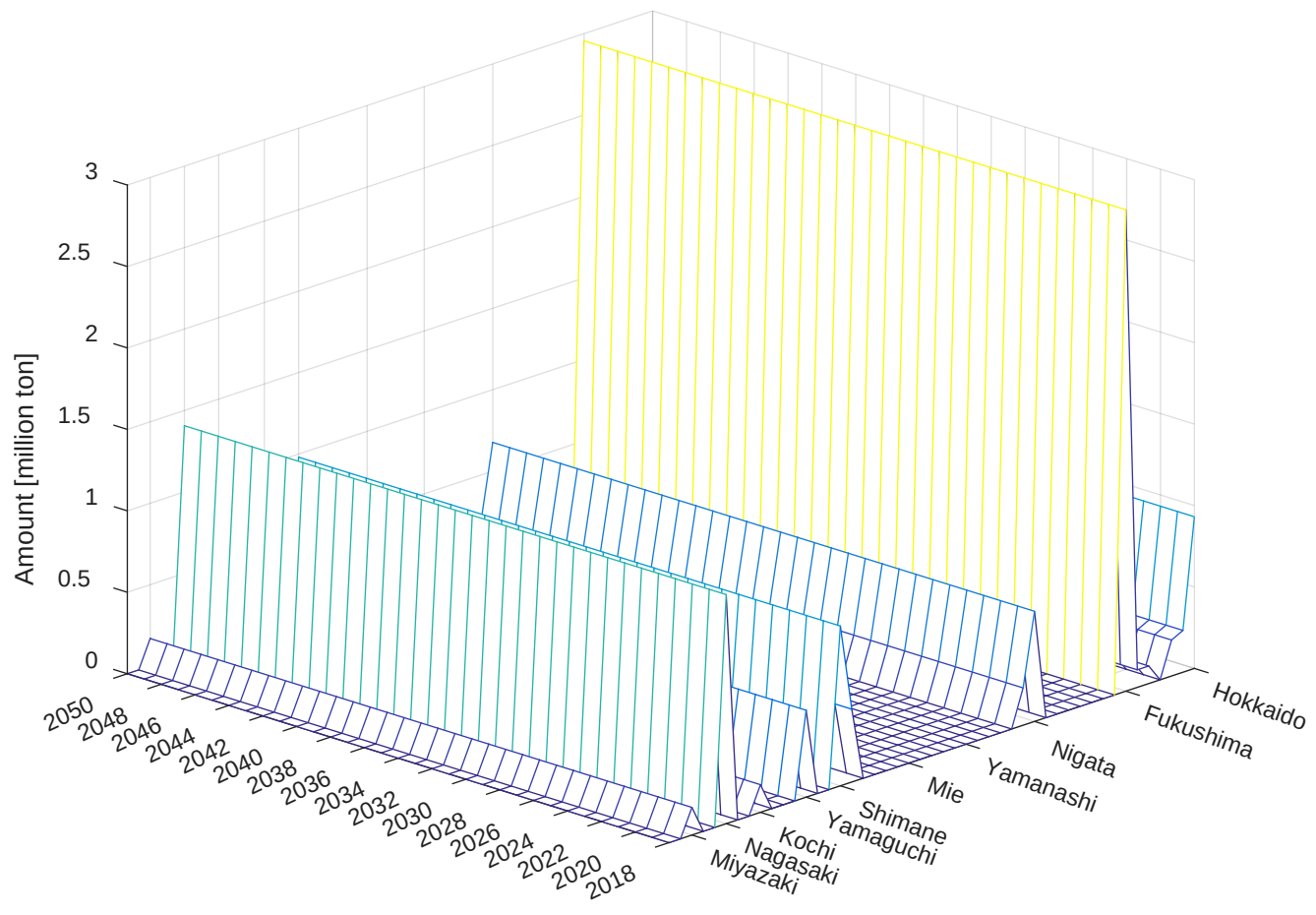

**Figure 4**. Yearly CCS deployment obtained from the proposed model between 2018–2050 for Scenario 1 enabling a 67% $CO_2$ reduction.

***Scenario 2:*** Due to the consideration of the power mix resilience constraint, the maximum amount of $CO_2$ which can be offset in this scenario is 59%, lower than that for Scenario 1 but still significantly higher than the 34% achieved in a similar scenario but without CCS in [13]. Details on the deployment of solar, wind, and CCS are shown in Figures 5–6, respectively.

As can be observed, solar deployment in this scenario begins earlier and is distributed more evenly over prefectures than was the case for Scenario 1. Moreover, the overall solar deployment level is increased, while wind is decreased because the power mix resilience constraint is taken into account. On the other hand, the CCS deployment level is not affected.

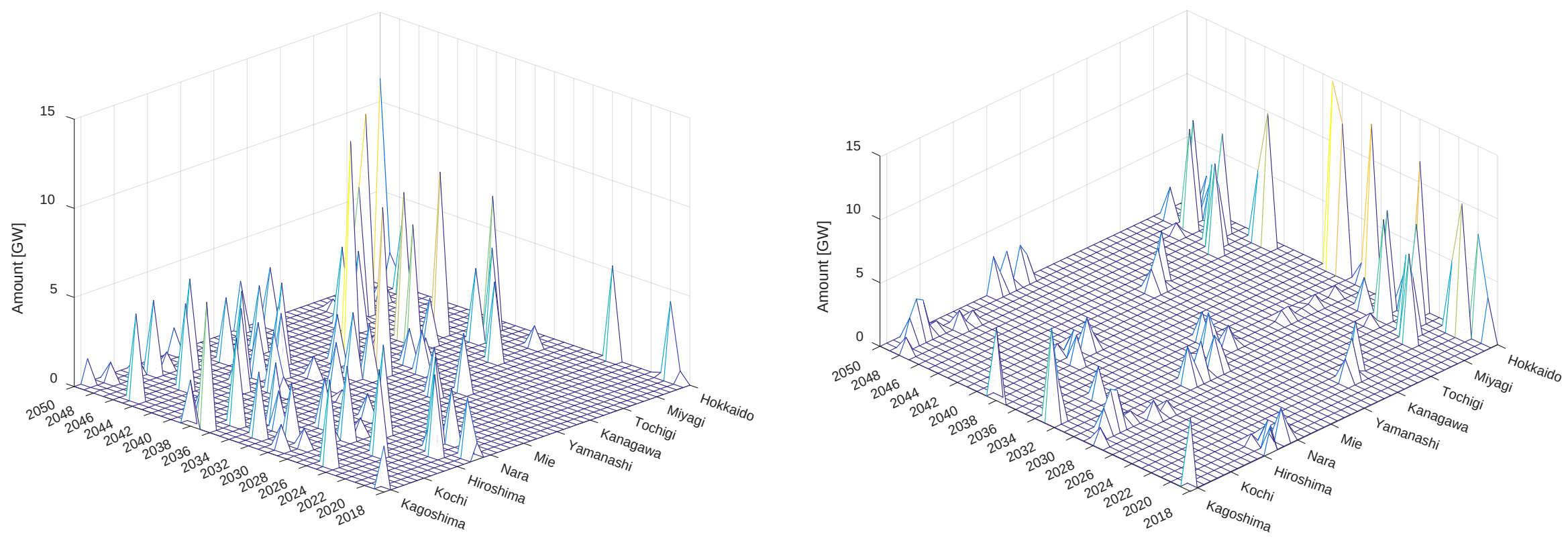

**Figure 5**. Yearly solar (left) and wind (right) deployment obtained from the proposed model between 2018–2050 for Scenario 2 enabling a 59% $CO_2$ reduction.

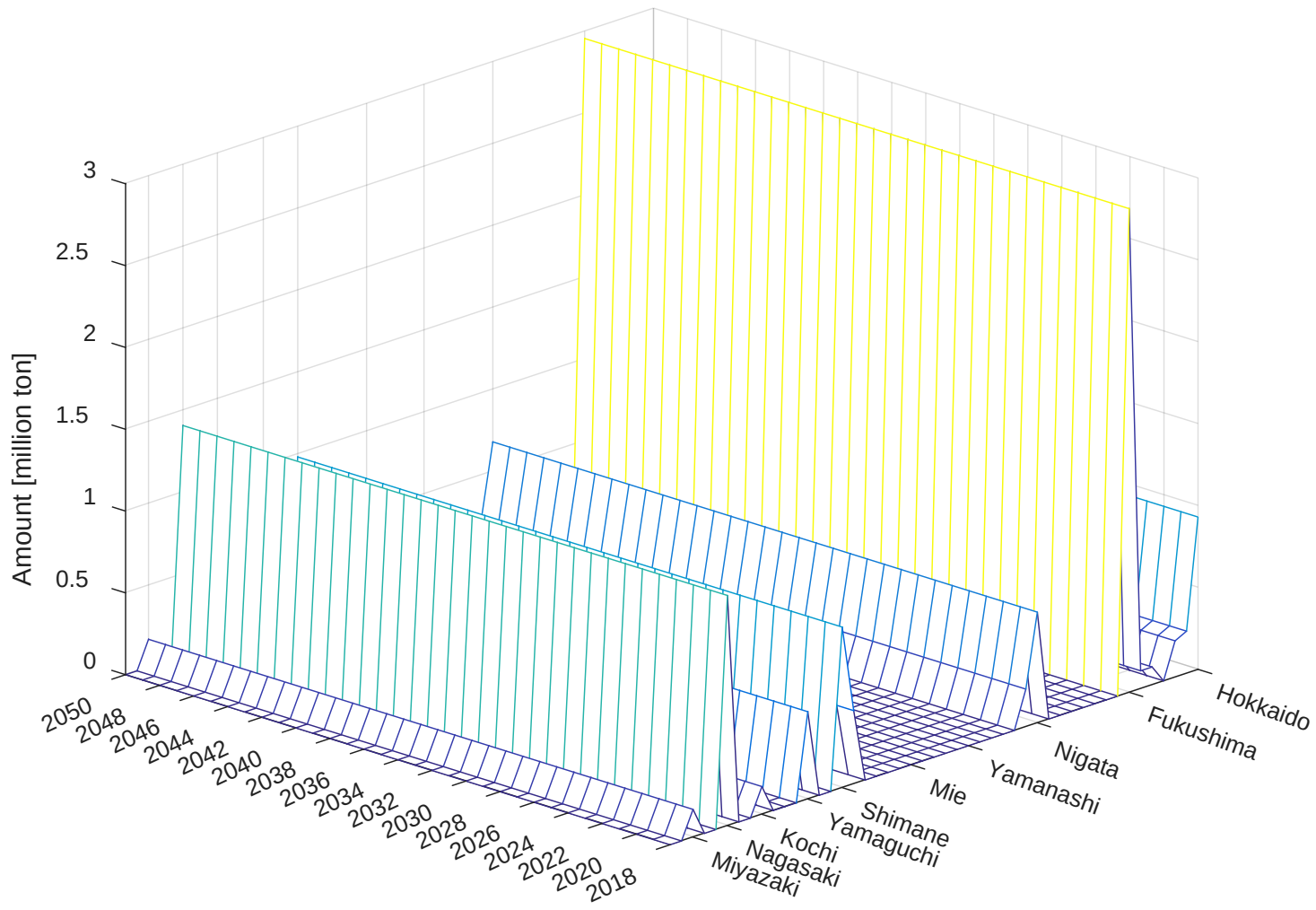

**Figure 6**. Yearly CCS deployment obtained from the proposed model between 2018–2050 for Scenario 2 enabling a 59% $CO_2$ reduction.

***Scenario 3:*** The maximum amount of $CO_2$ reduced in this scenario is 55%, with simulation results shown in Figures 7–8. As seen, both solar and wind installation start much later than in Scenarios 1 and 2. On the other hand, the yearly CCS deployment is much higher than that for Scenarios 1 and 2, but it is only deployed in the early years of the investigated period leading to a reduced overall contribution to $CO_2$ reductions.

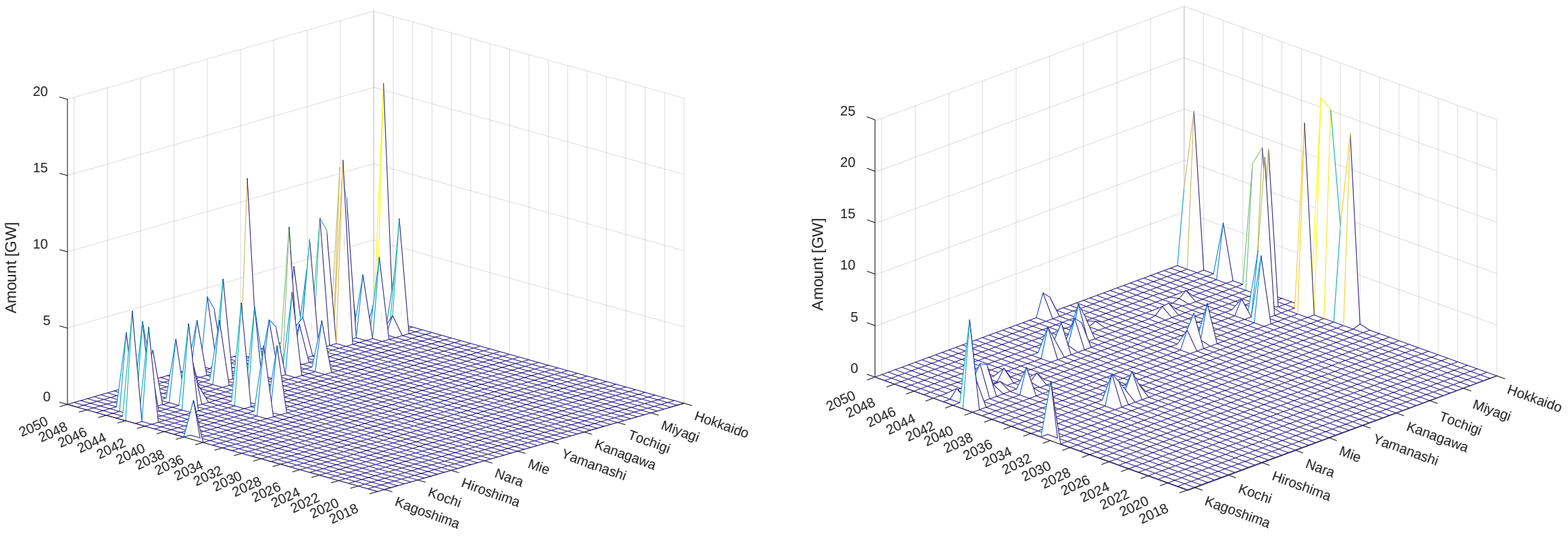

**Figure 7**. Yearly solar (left) and wind (right) deployment obtained from the proposed model between 2018–2050 for Scenario 3 enabling a 55% $CO_2$ reduction.

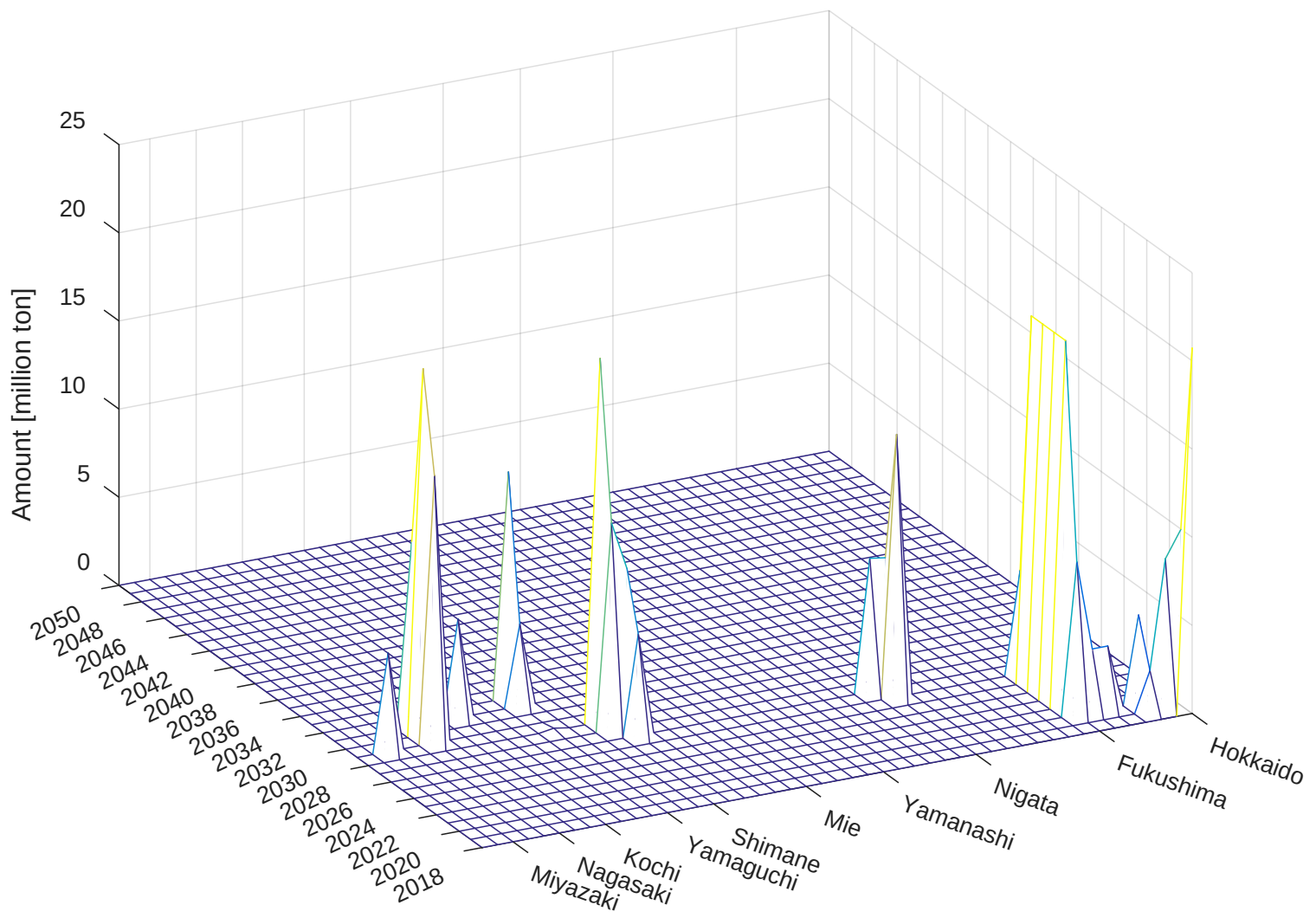

**Figure 8**. Yearly CCS deployment obtained from the proposed model between 2018–2050 for Scenario 3 (without equal CCS deployment) enabling a 55% $CO_2$ reduction.

***Scenario 4:*** The maximum amount of $CO_2$ can be offset in this scenario is 58.3%, substantially smaller than that for Scenario 1 when the CCS is evenly deployed over time, a similar observation with that of Scenario 3, demonstrating the benefit of constant, incremental deployment regimes. Details on deployment of solar, wind, and CCS are shown in Figures 9–10, respectively.

Similar to Scenario 2, the solar installation is more evenly distributed over prefectures in higher amounts, thanks to the existence of the power mix resilience constraint. CCS deployment is very different to that of Scenario 3. Specifically, CCS is deployed only in the first year of the considered period in all prefectures with physical storage capability, with a significant amount to be deployed in Fukushima prefecture, as detailed in Figure 10.

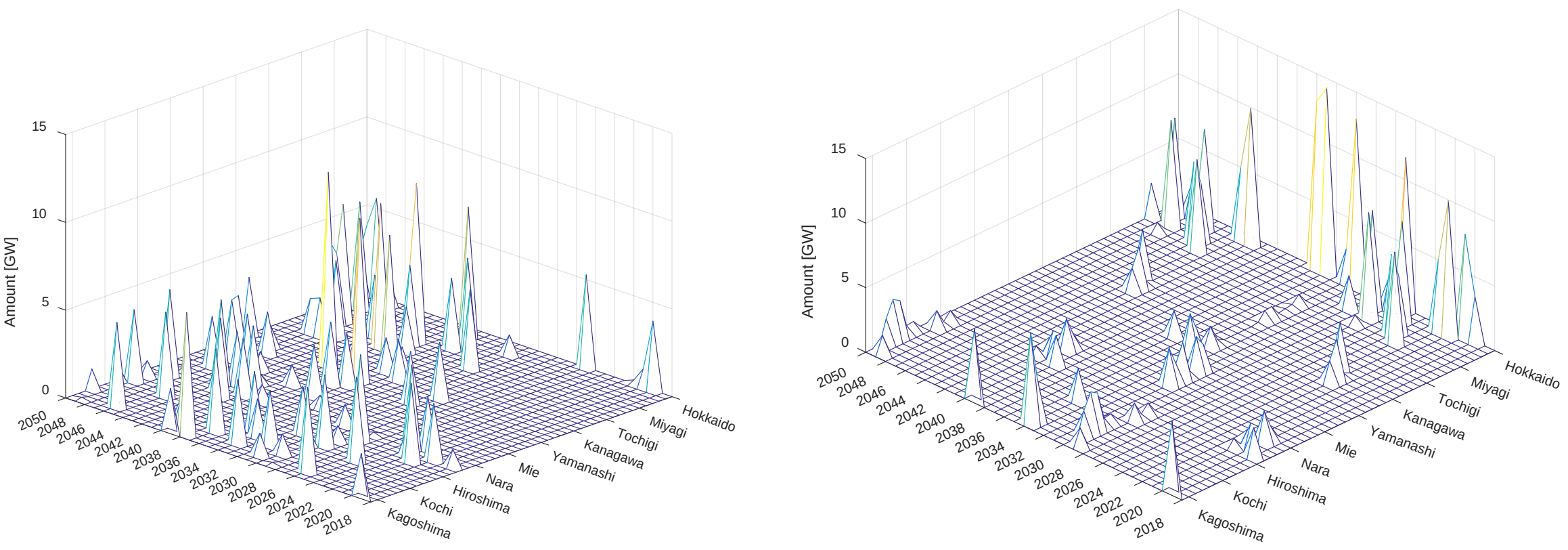

**Figure 9**. Yearly solar (left) and wind (right) deployment obtained from the proposed model between 2018–2050 for Scenario 4 enabling a 58.3% $CO_2$ reduction.

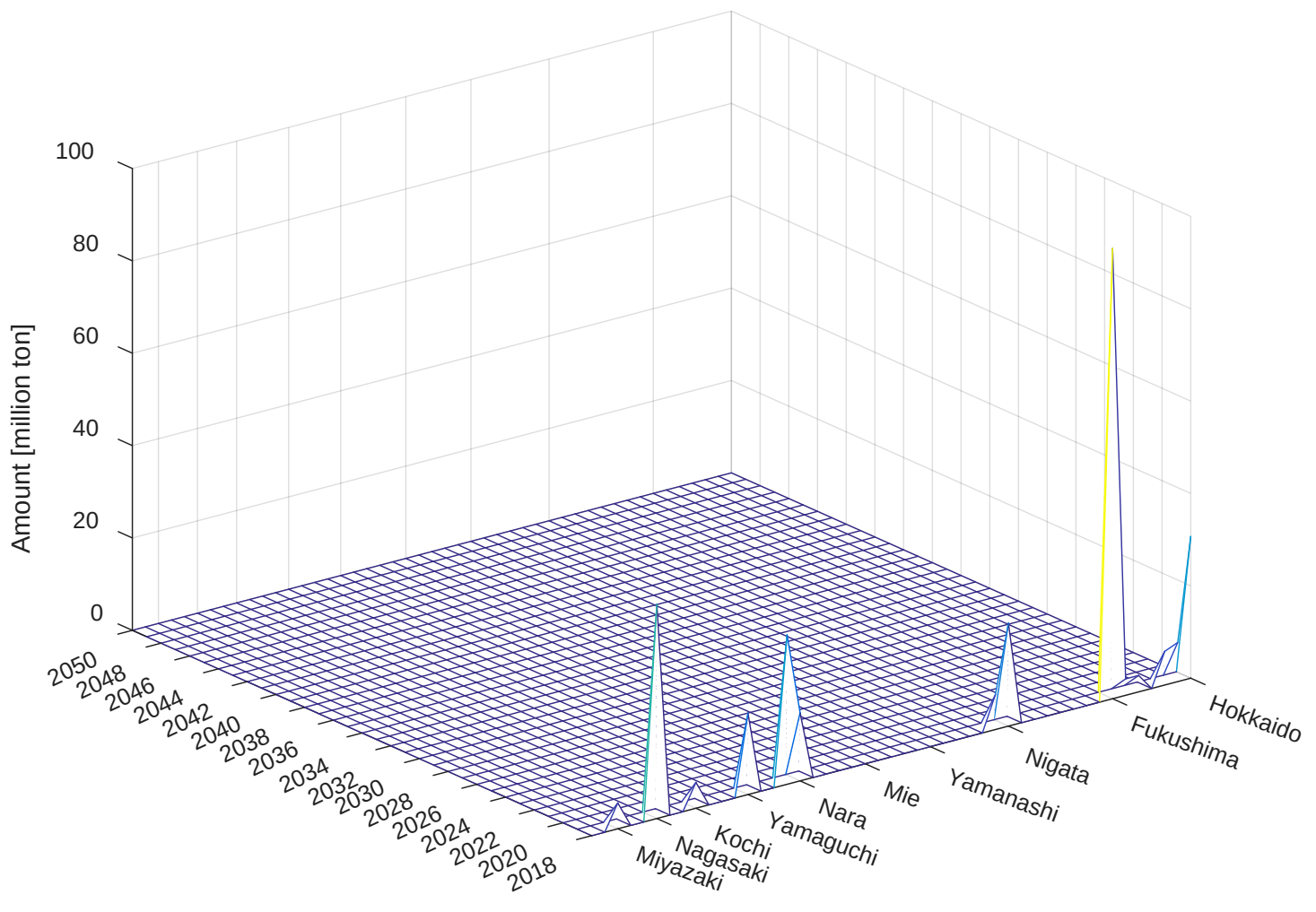

**Figure 10**. Yearly CCS deployment obtained from the proposed model between 2018–2050 for Scenario 4 enabling a 58.3% $CO_2$ reduction.

Additionally, we consistently observe through four considered scenarios that CCS deployment in Fukushima is at the largest amount compared to other prefectures. This is an interesting observation and provides important socioeconomic implications for developing appropriate policies to revitalize Fukushima after the Great East Earthquake in 2011.

In the following, we intend to analyze how the variation of uncertain factors, i.e., the carbon price $cp(t)$, the CCS unit price $ccsp(t)$, and the unit CCS transportation cost $gt(t)$, affect proposed model outcomes.

As mentioned before, no CCS trading occurs between prefectures in all four investigated scenarios. Hence, we aim to reduce the CCS unit price and the unit CCS transportation cost while increasing the carbon price to stimulate the potential for CCS trading. In spite of these modifications, CCS trading does not occur, suggesting that sequestration of local $CO_2$ may be preferable in all cases.

As an example, when the unit CCS transportation cost is set to zero, i.e., $gt(t) = 0$, the CCS unit price is reduced to $gt(t) = 1$ Y/ton, and the carbon price is increased to $cp(t) = 210{,}000$ Y/ton, the results of Scenario 1 are not changed. If the carbon price is further increased, then the maximum amount of $CO_2$ reduction is also increased. For instance, if the carbon price exceeds 220,000 Y/ton, then the maximum $CO_2$ reduction amount is significantly increased to 70.69% for Scenario 1. The deployment of solar, onshore wind, and CCS are also impacted significantly, as shown in Figures 11–12. Both solar and wind deployments are now significantly boosted, especially for wind in Hokkaido in the first year, whereas CCS deployment is halted for one year in the prefectures Hokkaido, Aomori, Iwate, Akita, and Yamagata.

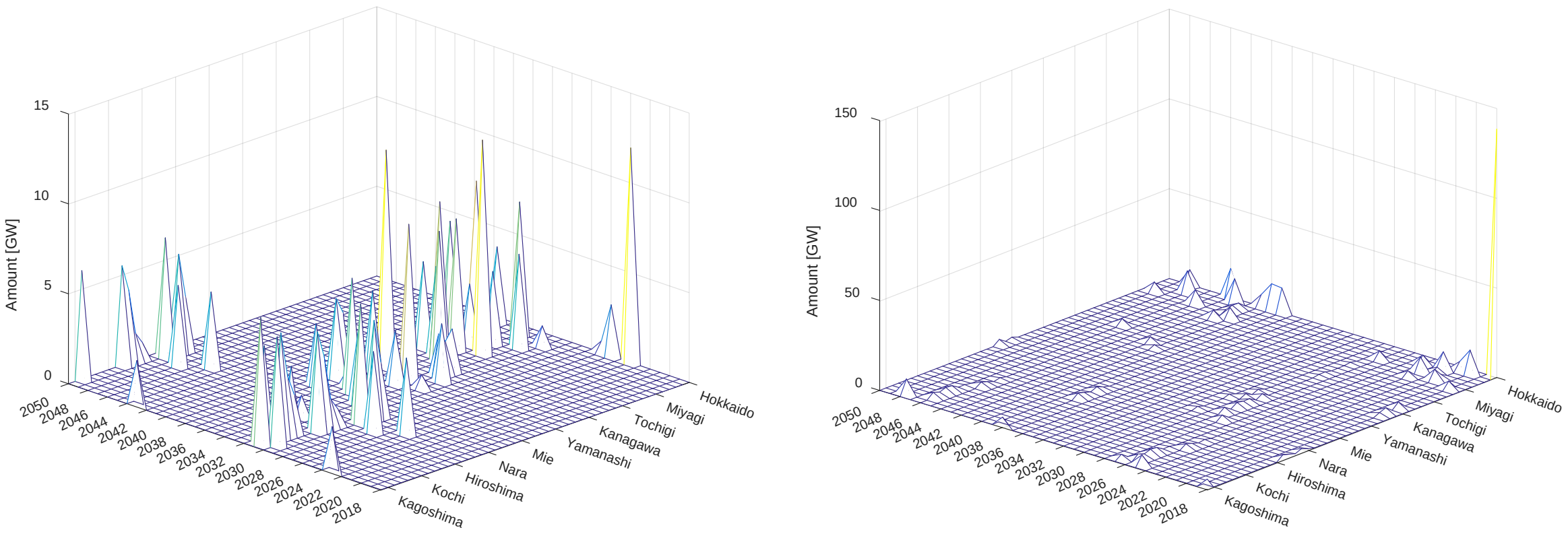

**Figure 11**. Yearly solar (left) and wind (right) deployment obtained from the proposed model between 2018–2050 for sensitivity analysis of Scenario 1, with a 70.69% $CO_2$ reduction.

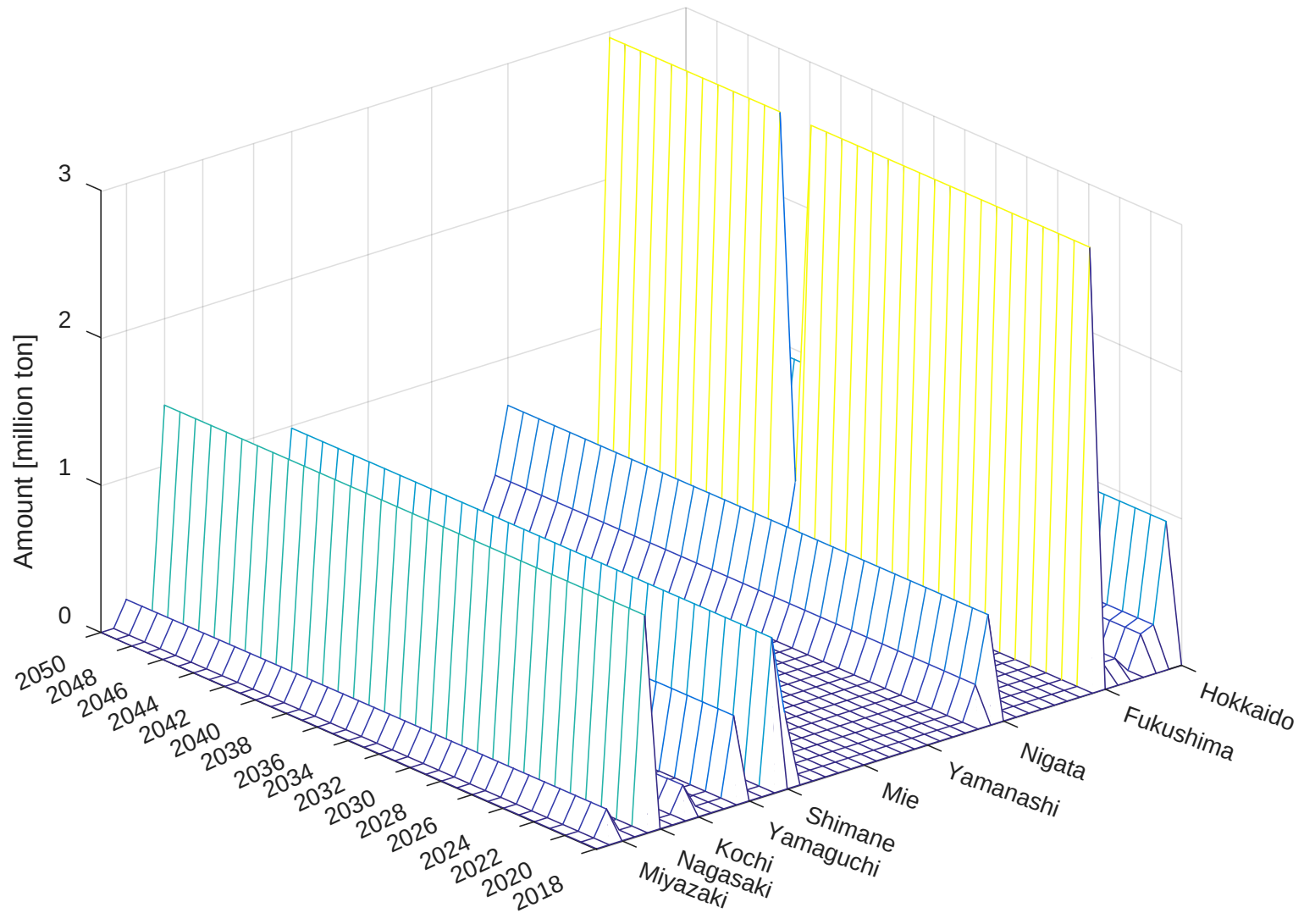

**Figure 12**. Yearly CCS deployment obtained from the proposed model between 2018–2050 for sensitivity analysis of Scenario 1, with a 70.69% $CO_2$ reduction.

## 5. Discussions

Here we discuss the environmental and economic performance of each carbon reducing technology under the prescribed scenarios, along with policy implications.

### 5.1 Environmental Analysis

Environmentally speaking, as was discussed in the results section, each scenario reduces $CO_2$ by between 55 and 67% of total $CO_2$ emissions in Japan, as shown in Figure 13.

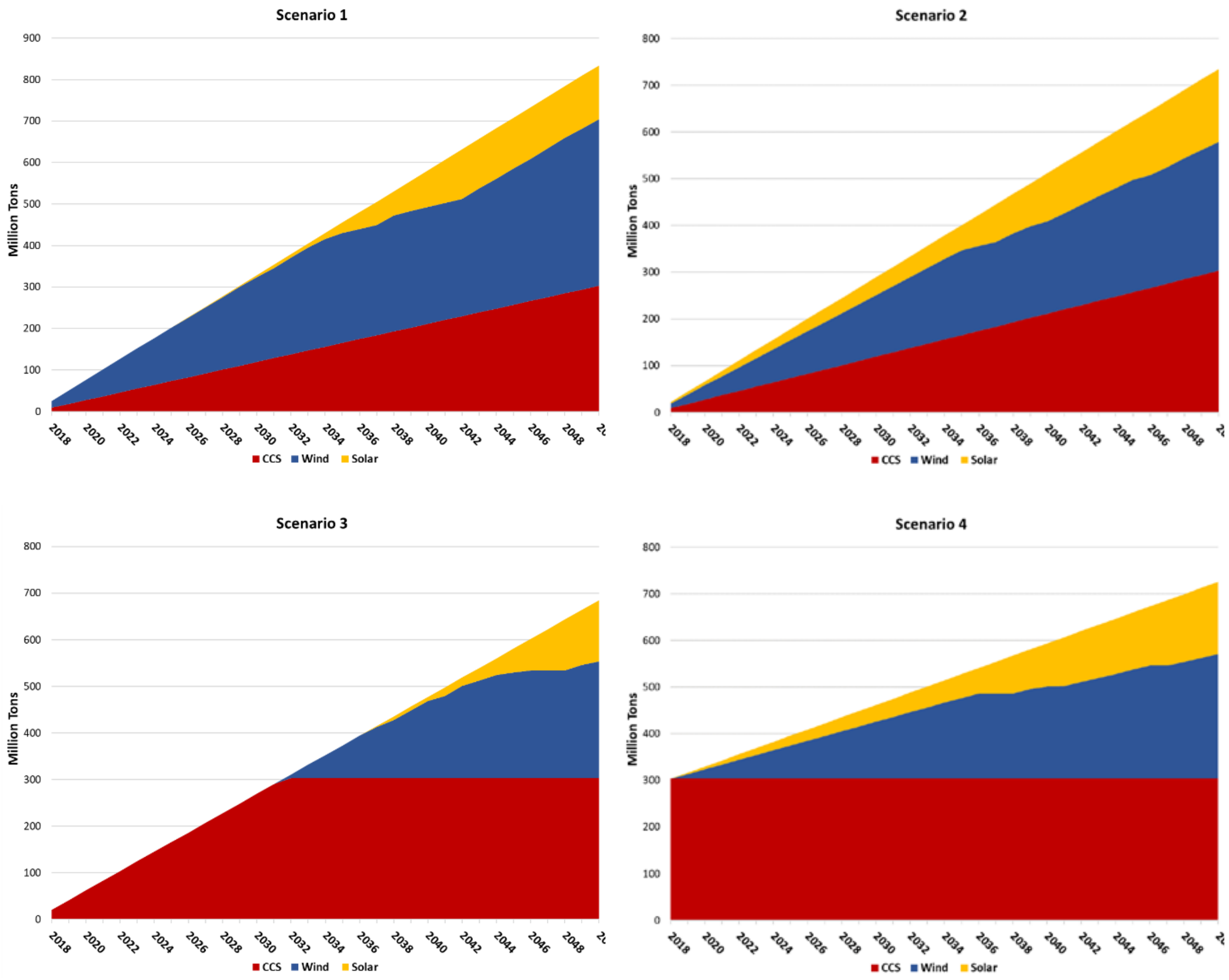

**Figure 13.** $CO_2$ emission reduction by technology between 2018 and 2050 in each scenario.

For all scenarios, CCS contribution to carbon emission reductions is restricted to a maximum deployment level of 303.3 million tons by 2050, with deployment regimes and timelines varied according to scenario settings.

In scenario 1 and scenario 2, where the physical limits of CCS are filled in a linear fashion over time, we observe an increasing 'wedge-shaped' contribution from CCS to overall emission reductions, of approximately 36.3% in scenario 1 and 41.3% in scenario 2, respectively. The difference in contribution from CCS toward overall emission reductions can be explained by an increased role for solar power in scenario 2, due to power mix resilience requirements, which both reduces wind power's contribution and the overall level of $CO_2$ reducing capability. Scenario 1 offsets the most $CO_2$ among all scenarios, however the $CO_2$ emission reducing technology diversity may be described as relatively poor, with wind having the largest role to play.

Scenario 3 rapidly deploys CCS to the year 2032, at which point new CCS deployment is ceased and wind begins to play a role. Solar is introduced beginning in the year 2037, finally contributing approximately 19.1% to $CO_2$ emission offsets by 2050, compared to 36.5% for wind, and the majority share of 44.3% for CCS. By introducing the resilience requirement in addition to scenario 3 assumptions in scenario 4, the maximum capacity of CCS is installed in the first year of simulation, with no additional deployments thereafter. Wind and solar are deployed throughout the simulated timeline, and as a result contribute approximately 21.3% and 36.9% to $CO_2$ offsets by 2050 compared to a 41.8% contribution from CCS.

Only in scenario 1, which enables the greatest $CO_2$ reduction impact is CCS not the majority source of $CO_2$ offsets. On the other hand, where CCS provides the highest portion of $CO_2$ offsets, in scenario 3, the lowest level of $CO_2$ reduction impact is realized.

### 5.2 Economic Analysis

In terms of economic outcomes, Figure 14 describes the investment costs, electricity generation and carbon tax revenues along with cumulative profitability (i.e., payback periods) considering technology deployment regimes for each scenario.

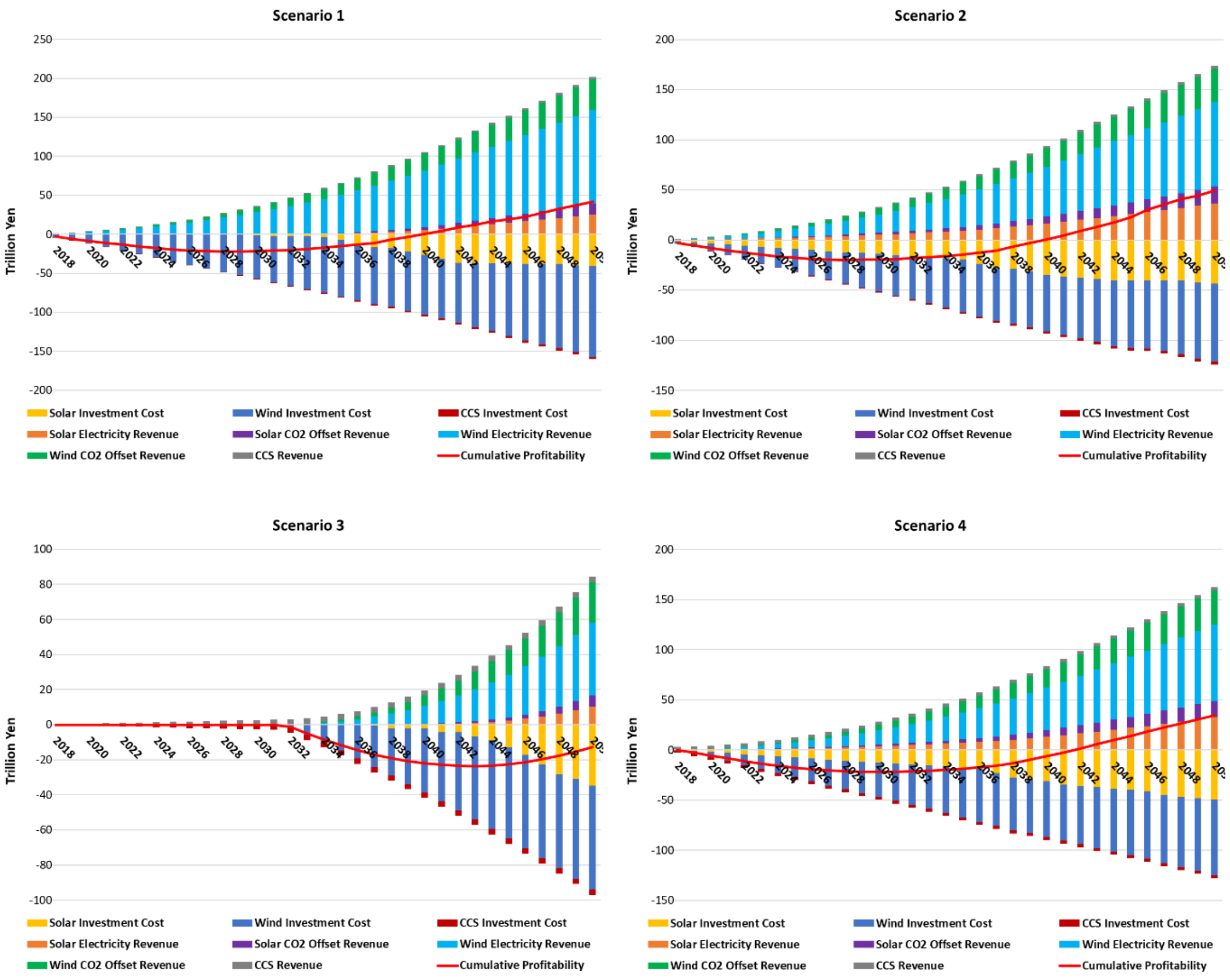

**Figure 14.** Investment costs and revenues per technology between 2018 and 2050 in each scenario

Scenario 1, which reduces the highest amount of $CO_2$ overall becomes profitable in the year 2040, representative of a 22-year payback period on the solar, wind and CCS investments. Wind power is responsible for the majority of revenues, yielded from electricity generation and carbon offsets (which are then sold as carbon credits at the prescribed carbon price). Cumulative profitability of scenario 1 peaks in the year 2050 at 41.6 trillion yen, which will continue to grow as long as investment in renewables is sustained into the future.

Scenario 2, by introducing resilience and therefore a large share of solar, sees revenues reduced for wind and increased for solar. Overall, the effect on investment payback period is negligible compared to scenario 1, with profitability also achieved in the year 2040. Although not as effective in reducing $CO_2$ as scenario 1, scenario 2 is more profitable by the year 2050, yielding a cumulative 49.3 trillion yen from electricity and carbon offset revenues. Deploying the lower cost technology of solar power ubiquitously yields financial benefits at the cost of some positive environmental impacts.

Scenario 3 defers the deployment of renewables to the year 2032 in preference for an exclusively CCS-based $CO_2$ offset regime. This has a stagnating effect on both investment and revenue streams, meaning that break-even is not achieved by 2050, although is likely to be achieved shortly thereafter. Alternatively, scenario 4, which locks in maximal CCS in the first year and deploys renewables consistently thereafter achieves profitability in the year 2042, and a final cumulative profit by the year 2050 of 34.8 trillion yen.

In terms of revenue generation, and appeal to investors, wind and solar power have an advantage over CCS, in that the benefits yielded are compounded over time, whereas CCS offsets $CO_2$ only once, when it is initially stored. On the other hand, considering the environmental efficacy of each technology, CCS has the advantage of massive storage capacity and consistently contributes a large portion of $CO_2$ offsets in each scenario.

### 5.3 Policy Implications

Japan, like other developed nations, faces the challenge of achieving carbon neutrality by the year 2050. In terms of the energy sector, Japan has significant solar and wind resources which can be tapped to contribute to its decarbonization, along with a moderate level of suitable CCS sites. The emergence of carbon pricing or a carbon tax, along with the continuation of the feed in tariff policy for large-scale renewables are likely to stimulate carbon offsetting mechanisms differently over time.

As was demonstrated by our scenario analysis, each technology has its own merits, ranging from a low cost or low barriers to deployment (i.e., solar), superior energy generation efficiency (i.e., wind), and superior carbon offsetting capacity in the investigated time frame (i.e., CCS). These merits will likely influence investors as to the timing and deployment regime engaged for this combination of key technologies. In terms of return on investment, our investigations also identified that waiting to deploy technologies, or using single technologies exclusively does not yield the best return on investment, and a combination of deployment of all available technologies while considering energy system resilience yields the best returns overall (scenario 2).

As has been the case to date, it is likely that feed in tariffs will gradually reduce over time, converging to between 8.5 and 11.5 yen per kilowatt hour ([36]; depending on region) after the initial feed in tariff contract period concludes (20 years for large scale solar and wind; [33]). On the other hand, as has been seen in other countries, carbon prices are likely to increase over time, as the need to achieve carbon neutrality becomes more urgent [37,38]. Our sensitivity analysis identifies an increased opportunity to reduce $CO_2$ via solar, wind and CCS if the carbon price increases as anticipated.

These changes to policy and economic stimuli are also likely to influence investment timelines. In terms of the cost of carbon reducing technologies, we anticipate that learning curves for renewables will continue to drive down prices, and this is likely to be the case for CCS too, but it may be too early to judge based on the limited success in commercialization to date. There is also an opportunity for the emergence of new technologies including direct air capture (DAC) that may play a similar or complementary role to CCS in the future [39]. A prudent approach may be to prioritize renewables in the short to mid-term, in anticipation of maturing and cheaper CCS approaches in the mid to long-term of decarbonization of the energy sector.

Finally, while wind and solar can be deployed in most prefectures of Japan, solar almost ubiquitously and wind along coastlines and in mountainous areas [28], CCS in our modeling is only considered for 14 prefectures (and their adjacent oceans; although additional prefectures such as Fukuoka and Ishikawa may have some potential) which have considerable storage capacities, as shown in Figure 15.

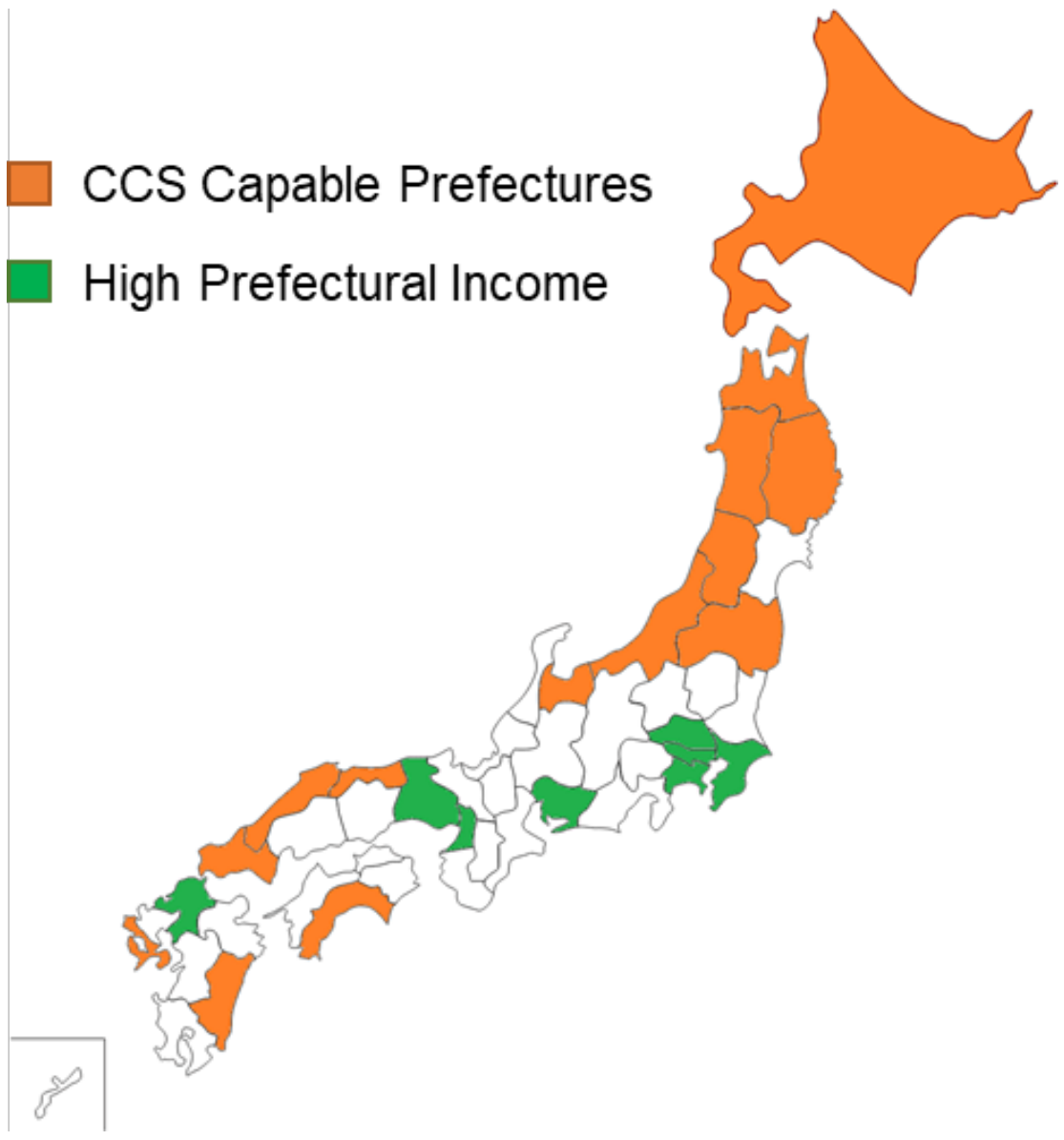

**Figure 15.** Location of Japanese prefectures with physical CCS storage sites and with high income.

All of the CCS capable prefectures are located in northern or western Japan, and none of them are recognized as 'high prefectural income' prefectures, with Hokkaido having the ninth highest prefectural income in Japan [40]. CCS provides the opportunity for a source of new income to prefectures which are not rich in traditional economic activity or have lower population densities. This may represent a way to reinvigorate prefectures which traditionally engaged in fossil fuel extraction activities (i.e., Hokkaido, Yamagata, Nagasaki, etc.), leading to new job opportunities and rural development. As shown through simulation results of all four scenarios in Section 4, Fukushima contributes the most and a substantially higher portion of CCS storage compared to other prefectures having physical CCS storage sites in our model. Thus, it can be a great opportunity for Fukushima to rebound its social and economic perspectives while being able to contribute to the whole national carbon emission reduction target.

## 6. Conclusions

This study proposes a linear programming model approach to assess the potential contributions toward carbon reduction of multiple technologies across a broad, diverse geographic area, to assess their economic and environmental merits. Taking into account various constraints reflecting the economically feasible deployment limits of renewable energy and CCS capacity as well as technological learning curves, carbon tax regimes and investment costs, the proposed approach is able to identify the maximum $CO_2$ offset achievable by 2050. Most importantly, this research contributes toward answering the critical question of where, when and how much of each technology is required to contribute to carbon neutrality goals, a critical issue around the globe by the year 2050.

Utilizing detailed data for the 47 prefectures of Japan as a case study, four scenarios were investigated including maximal deployment of technologies according to cost merits, and separately considering energy system resilience and various CCS deployment regimes. Although there are tradeoffs between scenarios in terms of the total $CO_2$ amount offset from the energy sector, and in terms of return of investment, the large potential contribution of CCS is recognized across scenarios. Further, the contributions of solar and wind for $CO_2$ offset capability and economic merit are also clarified, where solar is preferred in terms of a lower investment cost, and wind is recognized for its superior emission offset capability.

The findings of this research lead to important policy implications including the necessity for the continuation of the feed-in-tariff for large-scale renewables and the need for a carbon tax which increases over time as the need to reduce carbon emissions becomes more urgent. Also, in the case of Japan, the prefectures which benefit economically from having CCS storage capability are not those with high prefectural incomes. This finding identifies CCS as both an opportunity to contribute to reducing overall $CO_2$ emissions in Japan and also for creating jobs and aiding in the reinvigoration of rural areas both economically and in terms of associated social benefits. Although Japan is utilized as a case study nation in this study, the results can be applied to other nations, particularly those who share the geographic and limited fossil fuel resource challenges, and the need to diversify energy related environmental and economic returns.

This research has some limitations, including the fact that it only considers the energy sector, and does not contribute to the decarbonization of more difficult sectors such as industry and some industrial processes which should be incorporated into future studies.